\documentclass[prd,superscriptaddress,nofootinbib,floatfix,aps,preprint,numbers]{revtex4-1}

\makeatletter\providecommand\href@noop{\@secondoftwo}\makeatother

\newcommand{\gsim}{\gtrsim}
\usepackage{graphicx}
\usepackage{enumitem}
\usepackage{latexsym}
\usepackage{amsfonts}
\usepackage{amssymb}
\usepackage{xcolor}
\usepackage{CJK}
\usepackage[export]{adjustbox}
\usepackage{amsmath}
\usepackage{slashed}
\usepackage{dcolumn}
\usepackage{verbatim}
\usepackage{float}
\usepackage{multirow}
\usepackage{soul}
\usepackage[normalem]{ulem}
\usepackage{ulem}
\usepackage{hyperref}
\usepackage{tikz}
\usepackage{adjustbox}
\usepackage[compat=1.1.0]{tikz-feynman}
\usepackage{natbib}

\newcommand{\dd}{\mathrm{d}}

\newcommand{\Eprint}{\href}

\newcommand{\mpl}{{M_{\rm {pl}}}}
\tikzfeynmanset{
  fermion1/.style={
    /tikz/postaction={
      /tikz/decoration={
        markings,
        mark=at position 0.6 with {
          \arrow{>[length=5pt, width=4pt]};
        },
      },
      /tikz/decorate=true,
    },
  },
  fermion2/.style={
    /tikz/postaction={
      /tikz/decoration={
        markings,
        mark=at position 0.7 with {
          \arrow{>[length=4pt, width=4pt]};
        },
      },
      /tikz/decorate=true,
    },
  },
}

\allowdisplaybreaks

\begin{document}
\begin{CJK*}{UTF8}{gbsn}

\preprint{Imperial/TP/2020/MC/02}

\title{Neutrino-Assisted Early Dark Energy: Theory and Cosmology}

\author{Mariana Carrillo Gonz\'alez} \email{m.carrillo-gonzalez@imperial.ac.uk}
\affiliation{Center for Particle Cosmology, Department of Physics and Astronomy, University of Pennsylvania, Philadelphia, Pennsylvania 19104, USA}
\affiliation{Theoretical Physics, Blackett Laboratory, Imperial College, London, SW7 2AZ, U.K}
\author{Qiuyue Liang (梁秋月)} \email{qyliang@sas.upenn.edu}
\affiliation{Center for Particle Cosmology, Department of Physics and Astronomy, University of Pennsylvania, Philadelphia, Pennsylvania 19104, USA}
\author{Jeremy Sakstein} \email{sakstein@hawaii.edu}
\affiliation{Department of Physics \& Astronomy, University of Hawai'i, Watanabe Hall, 2505 Correa Road, Honolulu, HI, 96822, USA}

\author{Mark Trodden} \email{trodden@physics.upenn.edu}
\affiliation{Center for Particle Cosmology, Department of Physics and Astronomy, University of Pennsylvania, Philadelphia, Pennsylvania 19104, USA}

\date{\today}

\begin{abstract}
The tension between measurements of the Hubble constant obtained at different redshifts may provide a hint of new physics active in the relatively early universe, around the epoch of matter-radiation equality. A leading paradigm to resolve the tension is a period of early dark energy, in which a scalar field contributes a subdominant part of the energy budget of the universe at this time. This scenario faces significant fine-tuning problems which can be ameliorated by a non-trivial coupling of the scalar to the standard model neutrinos. These become non-relativistic close to the time of matter-radiation equality, resulting in an energy injection into the scalar that kick-starts the early dark energy phase, explaining its coincidence with this seemingly unrelated epoch. We present a minimal version of this neutrino-assisted early dark energy model, and perform a detailed analysis of its predictions and theoretical constraints. We consider both particle physics constraints --- that the model constitute a well-behaved effective field theory for which the quantum corrections are under control, so that the relevant predictions are within its regime of validity --- and the constraints provided by requiring a consistent cosmological evolution from early through to late times. Our work paves the way for testing this scenario using cosmological data sets.
\end{abstract}

\maketitle

\section{Introduction}

One of the more intriguing cosmological questions of recent years is raised by the discrepancy between measurements of the Hubble parameter obtained from observations of the cosmic microwave background (CMB) and higher values obtained using local distance indicators and strong lensing time-delays~\cite{Aylor:2018drw,Wong:2019kwg,Verde:2019ivm,Knox:2019rjx}\footnote{A recent reanalysis of the data \cite{Birrer:2020tax} has suggested that the errors on the H0LiCOW measurement may have been underestimated by a factor of three, and that accounting for this lowers the strong lensing measurement of the Hubble constant to $H_0=67.4_{-3.2}^{+4.1}$, bringing it into agreement with Planck.}. This so-called {\it Hubble tension} has stubbornly remained as datasets have been interrogated and updated, and measurements have become increasingly precise~\cite{Fitzpatrick:2000hh,Benedict:2006cp,Humphreys:2013eja,Efstathiou:2013via,Rigault:2014kaa,Becker:2015nya,Shanks:2018rka,Riess:2018kzi,Kenworthy:2019qwq,Spergel:2013rxa}. Naturally, theorists have wondered whether this discrepancy might be the first signal of new physics beyond the standard model of cosmology, and have pursued a number of possible explanations under this assumption~\cite{Lombriser:2019ahl,Vattis:2019efj,Pandey:2019plg,Agrawal:2019dlm,Desmond:2019ygn,Sakstein:2019qgn,Desmond:2020wep,Poulin:2018zxs,Bernal:2016gxb,DiValentino:2016hlg,Vagnozzi:2019ezj,Kreisch:2019yzn,Adhikari:2019fvb,Karwal:2016vyq,Mortsell:2018mfj,Poulin:2018cxd,Alexander:2019rsc,Niedermann:2019olb,Berghaus:2019cls}. 

Among the various early universe proposals, in which the pre-recombination cosmology is modified, in this paper we will be interested in early dark energy (EDE) models~\cite{Karwal:2016vyq,Mortsell:2018mfj,Poulin:2018cxd,Alexander:2019rsc,Niedermann:2019olb,Berghaus:2019cls}. In this scenario, a scalar field begins to contribute to the cosmic energy budget shortly before matter-radiation equality ($z_{\rm eq}\sim3000$) and rapidly becomes irrelevant thereafter. This causes the Hubble parameter to decrease more slowly than in the $\Lambda$CDM model, and thus reduces the sound horizon for acoustic waves in the photon-baryon fluid. As a result, the angular diameter distance to the surface of last scattering is also reduced, and the inferred Hubble constant is higher than in $\Lambda$CDM.

Simple models of early dark energy inevitably have a coincidence problem. The typical dynamics of a scalar field in the early universe is that the field remains frozen on its potential due to Hubble friction until the expansion rate drops below the mass of the field, at which point the field begins to roll. It is therefore necessary to arbitrarily choose the mass of the scalar in order to ensure that it begins to roll down its potential close to matter-radiation equality. Furthermore, beyond needing to make this tuning, the relevant mass, $m_\phi\sim 10^{-29}$ eV, is so small that it presents a technical naturalness challenge to the model. Quantum corrections to the mass will drive it to significantly higher values unless the EDE field is uncoupled from the rest of the matter in the universe, or the model possesses symmetries that protect the mass from radiative corrections. The latter approach has proved an attractive one for other areas in which a scalar field in invoked in cosmology, resulting in models of natural inflation \cite{Freese:1990rb} and natural quintessence \cite{Albrecht:2001xt}, based on axion-like fields with broken shift symmetries. However, when applied to early dark energy the predictions of the simplest models run afoul of supernovae constraints on the expansion rate~\cite{Poulin:2018dzj,Knox:2019rjx}, and more complicated models, while comprehensively solving this problem~\cite{Poulin:2018cxd,Smith:2019ihp,Capparelli:2019rtn}, introduce severe fine-tunings in their own right.

Two of us recently proposed a novel way to address these fine-tuning issues facing early dark energy models~\cite{Sakstein:2019fmf}. Our idea of {\it neutrino-assisted early dark energy} makes use of the coincidence between the upper limit on the sum of the neutrino masses ($\le 0.5$eV~\cite{Aghanim:2018eyx,Aker:2019uuj}) and the energy scale of matter-radiation equality (the neutrino temperature at $z=3000$ is $0.51$ eV.). The idea is that by conformally coupling the EDE field to neutrinos, we might allow much heavier EDE fields, which are not frozen on their potentials at equality, but instead receive an energy injection at this epoch as neutrinos transition from being relativistic to non-relativistic. Having been displaced by this ``kick", the EDE field then rolls down its potential and addresses the Hubble tension in much the same way as in regular EDE models. 

The essential physics behind this idea was laid out in~\cite{Sakstein:2019fmf}, and a simple toy model was presented. In this paper we perform a careful analysis of neutrino-assisted early dark energy. In the next section, we review the essential features of the model and describe how the basic physics works using a toy example, coupling an EDE field to a single massive neutrino. Next, we introduce a minimal model that constitutes a robust effective field theory that exhibits all of the qualitative features of the toy model. We carry out a careful treatment of the neutrino distribution function, and in particular pay extremely close attention to the problem of identifying reliable initial conditions for the field, finding that, in general, the model may exit the range of validity of the effective field theory at early times. In order to extract robust results, we make a simplifying assumption about the ultraviolet (UV) completion of the theory that picks out a subclass of theories for which the EFT remains a valid description during the entire evolution.

In Section~\ref{sec:quantum}, we analyze the quantum consistency and field theory constraints on models with this coupling. We demand that the predictions of the model not be unacceptably perturbed by quantum corrections, and that the model be radiatively stable. These considerations place a set of constraints on the parameters of the model, and in particular on the mass of the EDE field. Given the sizes of the couplings involved, we expect that these constraints will hold in a full theory coupled to the standard model. 

With a robust theory in hand, we then turn in Section~\ref{sec:cosmo} to a thorough analysis of the background cosmological evolution.  Using both analytic approximations and numerical solutions, we study how the effectiveness of the mechanism changes as the parameters of the model are varied, and also gain insights into details of how the model works, and how it leaves subtle imprints in the expansion history both before and after the epoch of recombination. A scan over these parameters identifies those regions in which neutrinos deliver the most pronounced kick to the early dark energy field. These are the regions in which the mechanism has the best chance of providing a viable solution to the Hubble tension, and which we suggest as the most promising starting points for detailed comparisons with cosmological datasets. Finally, in section \ref{sec:concs} we identify potential experimental and observational signatures that could constitute novel probes of the model. We briefly analyze model extensions where the EDE field can couple to other Standard Model particles and the dark sector, before concluding with a discussion of our results and directions for future work.

\section{Review of Neutrino-Assisted Early Dark Energy}
\label{sec:review}

\subsection{Overview of the mechanism}

In order to exemplify the essential physics of this mechanism, it is instructive, as in~\cite{Sakstein:2019fmf}, to strip away much of the structure of the standard model of particle physics, and to focus on a simplified setup consisting of a single neutrino species without gauge symmetries. While this will remain a useful description here, we will show in the next section that crucial theoretical and phenomenological improvements can be made by including a mass for the EDE field, and by generalizing the form of its coupling to neutrinos. 

To start with though, consider the action, 
\begin{equation}
\label{eq:act}
S=\int d^4 x\sqrt{-g}\left[\frac{\mpl^2}{2} R(g)-\frac12\nabla_\mu\phi\nabla^\mu\phi-V(\phi)\right]+S_{\tilde{\nu}}[\tilde g_{\mu\nu}],
\end{equation}
where $S_{\tilde{\nu}}$ is the action for the neutrino sector but with all contractions made with a conformal metric 
$\tilde{g}_{\mu\nu}=e^{2\beta\frac{\phi}{\mpl}}g_{\mu\nu}$ rather than with $g_{\mu\nu}$. Expanding out the conformal coupling, we may write this in the equivalent form \cite{Wetterich:2014bma,Burrage:2018dvt}
\begin{equation}
\label{eq:act2}
S=\int d^4 x\sqrt{-g}\left[\frac{\mpl^2}{2} R(g)-\frac12\nabla_\mu\phi\nabla^\mu\phi-V(\phi)+i\bar\nu\gamma^\mu\overset{\text{$\leftrightarrow$}}{\nabla}_\mu\nu-m_\nu\left(1+\beta\frac{\phi}{\mpl}+\cdots\right)\bar\nu\nu\right],
\end{equation}
where all contractions are now made with the metric $g_{\mu\nu}$, $\overset{\text{$\leftrightarrow$}}{\nabla}\equiv(\overset{\text{$\rightarrow$}}{\nabla}-\overset{\text{$\rightarrow$}}{\nabla})/2$, and we have performed a field redefinition of the neutrino field, $\nu= e^{3\beta \phi/ 2 \mpl} \tilde{\nu}$.

On a Friedmann-Robertson-Walker background in coordinate time, the scalar field equation of motion (EOM) is
\begin{equation}\label{eq:eom0}
\ddot{\phi} + 3H\dot{\phi} + \frac{\partial V_{\rm eff}}{\partial \phi} =0 ,
\end{equation}
where we have absorbed the coupling to neutrinos into an effective potential
\begin{equation}
\label{eq:effpot}
V_{\rm eff}(\phi)\equiv V(\phi)-\beta{\Theta}(\nu)\frac{\phi}{\mpl},
\end{equation}
with 
\begin{equation}
   { \Theta}(\nu) ={g}_{\mu\nu}{\Theta}(\nu)^{\mu\nu}\equiv-\frac{2}{\sqrt{-{g}}}{g}_{\mu\nu}\frac{\delta S_\nu[ g_{\mu\nu}]}{\delta {g}^{\mu\nu}}= -{\rho}_\nu+3{p}_\nu\,
\end{equation}
the trace of the neutrino energy-momentum tensor, and
\begin{equation}
{\rho}_{\nu}(x)=\frac{g_{\nu}}{2 \pi^{2}} {T}_{\nu}^{4} \int_{x}^{\infty} d u \frac{u^{2}\left(u^{2}-x^{2}\right)^{1 / 2}}{e^{u}+1},\,~~~~ {p}_{\nu}(x)=\frac{g_{\nu}}{6 \pi^{2}} {T}_{\nu}^{4} \int_{x}^{\infty} d u \frac{\left(u^{2}-x^{2}\right)^{3 / 2}}{e^{u}+1}
 \label{eq:fermieos},
\end{equation}
where $g_\nu$ is the neutrino degeneracy, ${x} = m_\nu/{T}_\nu$,  and ${T}_\nu$ is the neutrino temperature.
Note that the neutrino's pressure and density are not individually conserved due to the scalar-neutrino coupling. In particular, one has
\begin{align}\label{eq:rhoneucons}
    \dot\rho_\nu+3H(\rho_\nu+P_\nu)=-\frac{\beta\dot\phi}{\mpl}(3P_\nu-\rho_\nu).
\end{align}
The behavior of ${\Theta}(\nu)$ is central to the success of this model. Before analyzing this,
let us rewrite the equation of motion in terms of $N = \log a$ and multiply by $1/\mpl$ on both sides,
\begin{equation}\label{eq:eom1}
\frac{\phi''}{\mpl} + \left(\frac{ H'}{H} + 3\right) \frac{\phi'}{\mpl} + \frac{1}{H^2\mpl}\frac{\dd V(\phi)}{\dd\phi}=\beta\frac{{\Theta}(\nu)}{H^2\mpl^2} \ ,
\end{equation}
where a prime denotes a derivative with respect to $N$. The advantage of expressing the EOM in this coordinate is that one can see that the behavior of $\phi$ depends crucially on the combination ${\Theta}(\nu)/(H^2 \mpl^2)$. At early times, $x = m_\nu/T_\nu \ll 1$, so that $3{p}-{\rho}$ is proportional to ${x}^2$ and ${\Theta}(\nu) \sim m_\nu^2 T_\nu^2$. During the radiation-dominated era, we can relate the Hubble expansion to the neutrino temperature via $3H^2\mpl^2=\pi^2T_\gamma^4/30 g_\star(T_\gamma)$ with $T_\gamma=(11/4)^{1/3}T_\nu$. Thus, ${\Theta}(\nu) /H^2\mpl^2$ tends to zero at high temperatures. Furthermore, at late times, the neutrino is non-relativistic, and $3{p}_\nu-{\rho}_\nu \approx - {\rho}_\nu$, so that ${\Theta}(\nu)/ H^2\mpl^2 \sim {\rho}_\nu/\rho_{\rm tot}  \ll 1$. Therefore, ${\Theta}(\nu) /H^2\mpl^2$ will exhibit a ``kick" feature around $x\approx 1$, corresponding to $T_\nu \approx m_\nu$, which occurs around the epoch of matter-radiation equality. The ultimate effect of this coupling is then to drive a similar kick feature in the scalar's  density parameter  $\Omega_\phi\sim V(\phi)/H^2$ around this time, causing a sudden increase followed by a rapid decay. The scalar thus acts as a source of EDE without the need to fine-tune its parameters to achieve coincidence with matter-radiation equality. This is occurs naturally because the temperature is of order the neutrino mass.  

\begin{figure}[!tb]
   \centering
   \includegraphics[width=0.5\textwidth]{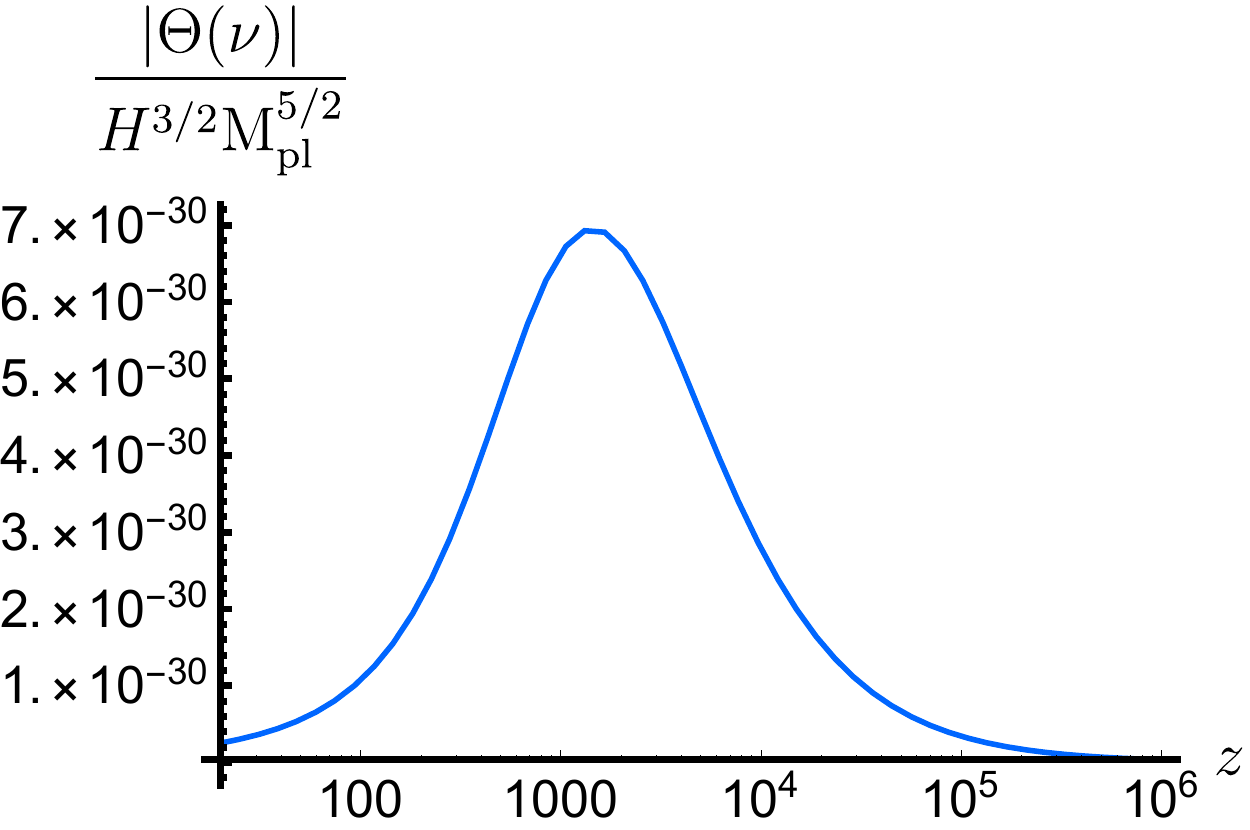}
 \caption{Evolution of the trace of the neutrino's energy-momentum tensor ${\Theta}(\nu)$ divided by $H^{\frac32}\mpl^{\frac52}$. From Equation~\eqref{eq:Omegaphi}, we can see that that this quantity drives the evolution of the EDE energy density parameter. As discussed in the text, this is responsible for the kick in $\Omega_\phi$.
 } 
 \label{fig,sketchOmega}
 \end{figure}

The mechanism is exemplified in Figure~\ref{fig,sketchOmega}. We have assumed that the EDE field remains at the minimum of the effective potential throughout its evolution process, and have chosen the EDE potential studied in \cite{Sakstein:2019fmf}, $V(\phi) = \frac{\lambda}{4} \phi^4$. The former assumption implies we can neglect the kinetic energy of the scalar and approximate $\Omega_\phi=V(\phi)/3H^2\mpl^2$.
With this choice of potential, the effective potential is minimized by
\begin{equation}\label{eq:phimin1}
 \phi_{\rm{min}} = - \left(\frac{\beta {\Theta}(\nu)}{\lambda  \mpl }\right)^{1/3},
\end{equation} 
so that the density parameter of the EDE scalar is
\begin{equation}\label{eq:Omegaphi}
 \Omega_\phi = \frac{1}{3H^2\mpl^2}\frac{\lambda}{4}\left(\frac{\beta {\Theta}(\nu)}{\lambda  \mpl }\right)^{4/3} = \frac{1}{12} \frac{\beta^{4/3}}{\lambda^{1/3}} \left(\frac{{\Theta}(\nu)}{H^{3/2}\mpl^{5/2}}\right)^{4/3} \ .
\end{equation}

While this simple toy model captures the general features of the mechanism, it relies on multiple theoretical assumptions that we have yet to substantiate. We need to ensure that $\beta\phi/\mpl \ll 1$, in order for our calculations to remain within the validity of the EFT. Similarly, we have neglected to include a mass for the scalar despite it being compatible with the symmetries of our action. We therefore need to explore the consequences of such an operator for our model, including the phenomenology and radiative stability.

\subsection{Minimal Model}

Motivated by the discussion above, we now move on to understanding the detailed constraints on the model arising from theoretical consistency and cosmological observations. We will work with the minimal model that is phenomenologically viable, and theoretically self-consistent. To that end, we consider a scalar coupled to a single massive neutrino species only, remaining agnostic as to how the neutrino mass term arose, and to whether the neutrinos are Dirac or Majorana. We will comment on possible generalizations of this model, including coupling to additional species, and other particles 
in Section \ref{sec:concs}. With this in mind, we consider the following generalized form of Equation~\eqref{eq:act}:
\begin{align}
\label{eq:act3}
S&=\int d^4 x\sqrt{-g}\left[\frac{\mpl^2}{2} R(g)-\frac12\nabla_\mu\phi\nabla^\mu\phi-\frac12m^2\phi^2-\frac{\lambda}{4}\phi^4\right]+S_{\tilde{\nu}}[A^2(\phi) g_{\mu\nu}]\\\nonumber&=\int d^4 x\sqrt{-g}\left[\frac{\mpl^2}{2} R(g)-\frac12\nabla_\mu\phi\nabla^\mu\phi-\frac12m^2\phi^2-\frac{\lambda}{4}\phi^4+i\bar\nu\gamma^\mu\overset{\text{$\leftrightarrow$}}{\nabla}_\mu\nu-m_\nu A(\phi)\bar\nu\nu\right],
\end{align}
where we have again redefined the neutrino field as ${\nu}=A^{3/2}(\phi)\tilde{\nu}$. In particular, we have included a mass term in the the scalar potential and have generalized the coupling function from exponential to consider general cases. Our motivation behind the former choice is effective field theory. This is the most general potential with a $\mathbb{Z}_2$ symmetry up to mass dimension four terms. Note that we have not imposed any constraints on the conformal coupling $A(\phi)$. However, in the following we will analyze Planck suppressed couplings which lead to a soft breaking of the $\mathbb{Z}_2$ symmetry. While terms such as $\phi^3$, which break the $\mathbb{Z}_2$ symmetry, can be radiatively generated, they will always be suppressed by factors of the Planck mass and can therefore be neglected in the self-interaction terms in the potential. The $\phi^4$ term provides an excellent fit to cosmological data when the mass is neglected \cite{Agrawal:2019lmo}. Adding a mass term leads to severe constraints from supernovae measurements of the late-time cosmic expansion rate but, from an EFT perspective, setting it to zero is unjustified. Indeed, if one imagines that this is a low energy EFT that arises from integrating out heavy degrees of freedom, then all operators compatible with the $\mathbb{Z}_2$ symmetry are expected to be present. Furthermore, once present, this mass is subject to quantum corrections from neutrino loops that arise from operators generated by the conformal coupling. It is presently unknown how large a mass is allowed by the data, and so a question of key importance in this work will be how small a choice we may make for the mass without spoiling the radiative stability of the model.

The generalization of the conformal coupling is, similarly, necessitated by EFT considerations. This follows from the behaviour of our model at early times derived in the previous subsection. Note that at early times we expect the $\phi^4$ term in the potential to dominate over the mass term since $|\phi_{\rm min}|$ is an increasing function of redshift, so we can neglect the mass term for the purpose of this discussion. If one takes $A(\phi)\sim\exp(\beta\phi/\mpl)$ then the effective potential at early times is
\begin{equation}
    V_{\rm eff}(\phi)=V(\phi)-\frac{\beta \phi}{\mpl}(3{P}_\nu-{\rho}_\nu),
\end{equation}
which is minimized by
\begin{equation}\label{eq:phimagmin}
    |\phi_{\rm min}|=\left[\frac{\beta}{\lambda\mpl}\left(3{P}_\nu-{\rho}_\nu\right)\right]^{\frac{1}{3}}.
\end{equation}
The reason that an exponential coupling will ultimately run afoul of EFT considerations is that this quantity grows with redshift. This is perhaps unexpected because $3{P}_\nu-{\rho}_\nu\rightarrow0$ for massless particles, and neutrinos become more relativistic in the past, but, since there is an exact cancellation between $3{P}$ and ${\rho}$ rather than a suppression by $(3{P}_\nu-{\rho}_\nu)/{\rho}_\nu$, the leading-order behavior scales like $|3{P}_\nu-{\rho}_\nu|\sim  m_\nu^2 T_\nu^2\sim (1+z)^2$ 
, which grows at early times. This implies that $|\beta\phi_{\rm min}/\mpl|>1$ at some early time (the specific time depends on the parameters of course). Since the neutrino coupling is a power series in $\phi/\mathcal{M}$ with $\mathcal{M}=\mpl/\beta$, $|\beta\phi_{\rm min}/\mpl|>1$ signifies that the field has exited the regime of validity of the EFT, implying that the model requires a UV-completion whose details become important around the redshift where this occurs.

Generalizing the coupling allows us to avoid this issue if we make one single assumption about the UV-completion in the scalar-neutrino sector, inspired by the strong coupling limit of string theory \cite{Damour:1994zq,Gasperini:2001pc,Brax:2010gi}. The assumption is that when the series is resummed, the coupling function has a minimum at some $\phi=-\bar\phi$ ($\bar\phi>0$). 
The effective potential is now 
\begin{equation}
\label{eq, vefflogA}
    V_{\rm eff}(\phi)=V(\phi)+(\rho_\nu-3P_\nu)\ln[A(\phi)].
\end{equation}
When $(\rho_\nu-3P_\nu)\dd\ln A/\dd\phi \gg V'(\phi)$ i.e. at high redshifts, one can ignore the potential and the field minimizes the coupling function solely i.e. $\phi_{\rm min}\approx -\bar\phi$. As the neutrinos redshift, the potential becomes increasingly important and the minimum of the effective potential differs from $\bar\phi$. Since $|\phi_{\rm min}|\le\bar\phi$, taking $\beta\bar\phi/\mpl<1$ is sufficient to ensure the evolution never leaves the range of validity of the EFT. A sketch of the difference between the cosmological evolution of the scalar with an exponential and generalized coupling is shown in Figure~\ref{fig,sketch}. We parameterize our ignorance of the UV-completion by taking
\begin{equation}
\label{eq,Acoupling}
    A(\phi)=1-\frac{A_2}{2}\bar\phi^2+\frac{A_2}{2}\left(\phi+\bar\phi\right)^2+\cdots \ ,
\end{equation}
where the constant term ensures $A(\phi)=1+\beta\phi/\mpl+\cdots$. (One does not need to make this choice but it can always be absorbed into constant redefinitions of the coordinates so we make it here for simplicity). Expanding this out, one finds
 \begin{equation}
\label{eq,effA}
    A(\phi)= 1 +\frac{\beta\phi}{\mpl}+\frac{\beta}{2\mpl}\frac{\phi^2}{\bar\phi} \ ,
\end{equation}
where we have identified $\beta=A_2 \mpl \bar{\phi}$. We neglect the $\cdots$ terms in Equation~\eqref{eq,Acoupling} since $\phi_{\rm min}\le\bar\phi$ for the entire cosmological evolution of the scalar, implying that the region of field space where these are relevant is never accessed.

     \begin{figure}
   \centering
   \includegraphics[width=0.5\textwidth]{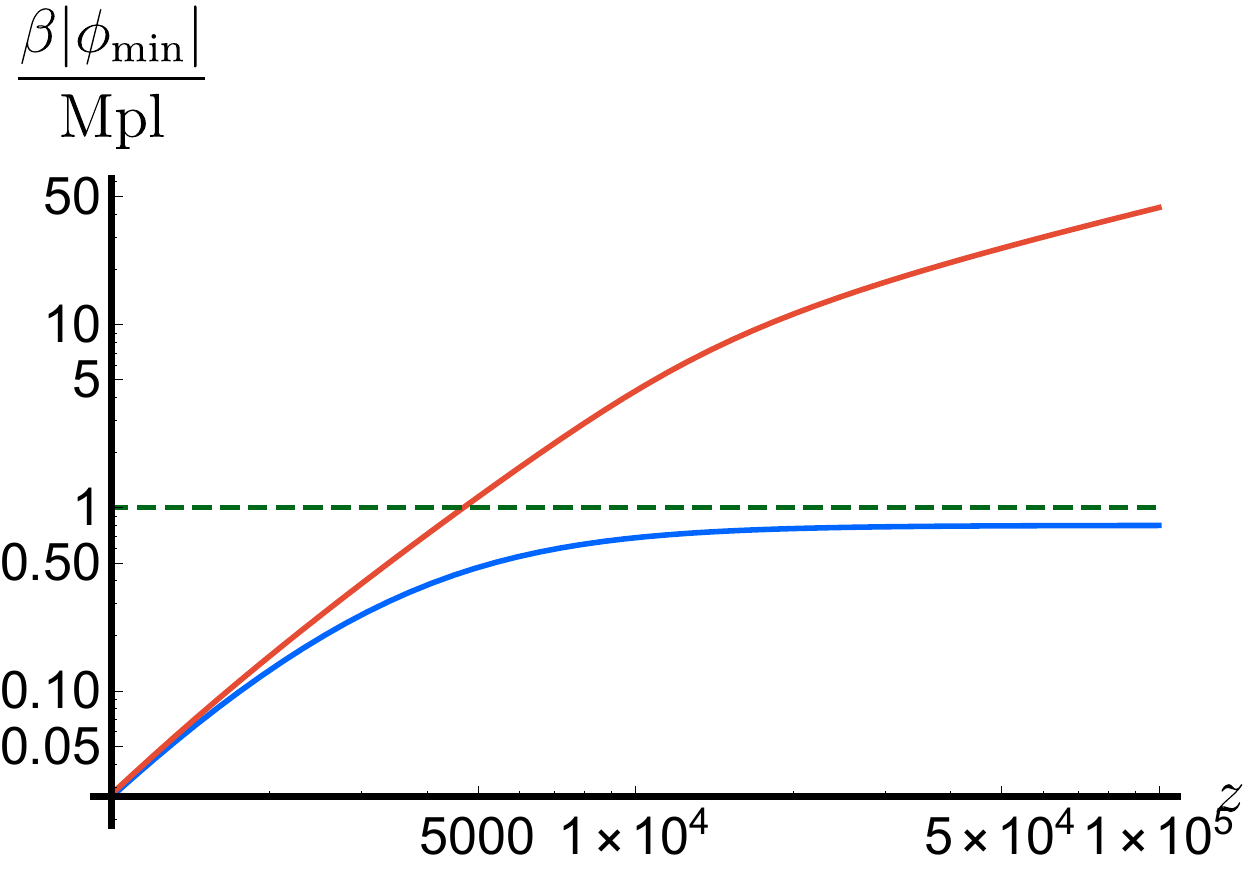}
 \caption{Sketch of the evolution of the EDE field at early times. The red line corresponds to the exponential coupling, leading to $\beta |\phi|/\mpl >1 $ at high redshift. The blue line shows the behavior of the field $\phi$ with a coupling given by Equation~\eqref{eq,effA}, with $\bar \phi = 0.8 \mpl/\beta$.} \label{fig,sketch}
 \end{figure}

\section{Quantum Corrections and Radiative Stability}
\label{sec:quantum}

In the previous discussion, we have explained why, on EFT grounds, we should include a mass term in our potential. Once this term is present, it can be troublesome in two ways. Firstly, the mass of the EDE field cannot be too large since it will contribute as an additional dark matter component which is tightly constrained by observations. Secondly, this mass should be radiatively stable to quantum corrections that arise from the scalar-neutrino coupling. To investigate this, we first expand the EDE field around the background value $\phi = \phi_{\rm min} + \varphi$. As discussed before, at early times, $ \phi_{\rm min} = -\bar\phi$ and at late times $\phi_{\rm min}$ deviates from $\bar \phi$. Expanding Equation~\eqref{eq:act3} and using Equation~\eqref{eq,effA}  we find that the resulting bare mass for perturbations of the EDE field is
\begin{equation}
    m_{\rm bare}^2=m^2+3\lambda \phi_{\rm min}^2 \ , \label{baremass}
    \end{equation}
and the coupling to neutrinos now reads
\begin{equation}
 \left[ \frac{\beta}{\mpl} \left(1+ \frac{\phi_{\rm min}}{\bar \phi} \right) \varphi +\frac{\beta }{2\mpl \bar \phi} \varphi^2  \right] m_\nu \bar \nu \nu .
\end{equation}
Furthermore, we now have both cubic and quartic self-interactions of the field $\varphi$. 

We now proceed to analyze the 1-loop quantum corrections to the EDE mass due to these couplings. We first consider the corrections arising from self interactions of $\varphi$. At 1-loop, these come from the diagrams in Figures~\ref{fig,loop1}(b) and \ref{fig,loop1}(c). The graph in Figure~\ref{fig,loop1}(b) leads to corrections to the mass-squared of order $ \lambda^2 \phi_{\rm min}^2   \ , $ while the one from Figure~\ref{fig,loop1}(c) gives corrections of order $ \lambda m_{\rm bare}^2   \ . $ Provided that the quartic self-coupling $\lambda$ is less than unity, the corrections from $\varphi$ loops will be smaller than the bare mass by a factor of $\lambda$, and hence will never significantly change the classical value. While this is clear for the correction from Figure~\ref{fig,loop1}(c), for the correction from Figure~\ref{fig,loop1}(b), we must analyze two different cases. If the term $m^2$ is equal or larger than the $\lambda \phi_{\rm min}^2 $ term in Equation~\eqref{baremass}, then $  m_{\rm bare}^2\simeq m^2\geq\lambda \phi_{\rm min}^2 >\lambda^2\phi_{\rm min}^2 $ holds. If, on the other hand, the $\lambda\phi_{\rm min}^2 $ term dominates, we have  $  m_{\rm bare}^2\simeq\lambda \phi_{\rm min}^2 >\lambda^2 \phi_{\rm min}^2 $. In either case, the correction is negligible.

Now consider corrections arising from the coupling to neutrinos. In principle, the quadratic coupling may contribute to the quantum corrections via a neutrino loop as in Figure~\ref{fig,loop1}(d) if the neutrinos are Majorana. This diagram does not exist for Dirac neutrinos, so from here on we restrict our model to this case, and focus on the quantum corrections arising from the linear coupling given by the diagram shown in Figure~\ref{fig,loop1}(a). Note that the linear coupling vanishes at early times, but at late times $|\phi_{\rm min}| \ll \bar \phi$ so the coupling constant multiplying $\varphi \bar \nu \nu$ tends to $\beta m_\nu /\mpl$. This gives a quantum correction to the mass term of the form
\begin{equation}
    \delta m^2 = \frac{\beta^2}{4\pi^2}\left(\frac{m_{\nu}}{\mpl}\right)^2m_\nu^2 \ . \label{qmass}
\end{equation}
In order to have a radiatively stable theory; that is, one in which the loop corrections are smaller than the tree level result, we therefore require that
\begin{equation} \label{mass}
    m^2+12\lambda\phi_{\rm min}^2 \gsim \frac{\beta^2}{4\pi^2}\left(\frac{m_{\nu}}{\mpl}\right)^2m_\nu^2 \ .
\end{equation}

\begin{figure}
  \centering
  \begin{tikzpicture}
  \begin{feynman}
   \vertex (a1);
    \vertex[right=1.cm of a1] (a2);
    \vertex[right=0.5cm of a2] (c2);
    \vertex[below =1. cm of c2] (c3){(a)};
     \vertex[right=1cm of a2] (a3);
     \vertex[right=1 cm of a3] (a4);
     
     \vertex[right=1cm of a4] (a5);
    \vertex[right =1 cm of a5] (a6);
    \vertex[right=0.5cm of a6](c4);
    \vertex[below =1 cm of c4] (c5){(b)};
     \vertex[right=1.cm of a6] (a7);
     \vertex[right=1. cm of a7] (a8);
     
     \vertex[right=1.cm of a8] (a9);
     \vertex[right=1cm of a9] (a10);
     \vertex[above=1cm of a10] (b1);
     \vertex[below =1. cm of a10] (c6){(c)};
     \vertex[right=1 cm of a10] (a11);
     
      \vertex[right=1.cm of a11] (a12);
    \vertex[right =1. cm of a12] (a13);
    \vertex[above=1cm of a13] (b2);
    \vertex[below =1. cm of a13] (c7){(d)};
     \vertex[right=1.cm of a13] (a14);

        \diagram* {
      (a1)-- [dashed, edge label' = $\varphi$](a2)--[fermion1, half left,edge label = $\nu$](a3)-- [dashed,edge label' = $\varphi$](a4) ,
      (a3)--[fermion1, half left,edge label = $\bar \nu$](a2),
        };
         \diagram* {
      (a5)-- [dashed, edge label' = $\varphi$](a6)--[dashed, half left,edge label = $\varphi$](a7)-- [dashed,edge label' = $\varphi$](a8) ,
      (a6)--[dashed, half right,edge label' = $\varphi$](a7),
        };
         \diagram* {
      (a9)-- [dashed, edge label' = $\varphi$](a10)-- [dashed,edge label' = $\varphi$](a11) ,
      (a10)--[dashed,  half right,edge label' =  $\varphi$](b1),
      (a10)--[dashed,  half left,edge label = $\varphi$ ](b1),
        };
         \diagram* {
      (a12)-- [dashed, edge label' = $\varphi$](a13)-- [dashed,edge label' = $\varphi$](a14) ,
      (a13)--[fermion2, half left,edge label = $\bar \nu$](b2),
      (a13)--[solid, half right,edge label' = $ \nu$](b2),
        };
  \end{feynman}
\end{tikzpicture} 
 \caption{One-loop diagrams that contribute to the quantum corrections of the EDE field's mass. The dashed line represents the early dark energy field $\phi$, and the solid line represents the neutrino $\nu$ and $\bar \nu$. The interaction vertex between $\phi$ and the neutrino comes from the effective potential in Equation~\eqref{eq, vefflogA}, and the effective coupling Equation~\eqref{eq,effA}. The final diagram (d) is only present for Majorana neutrinos, as discussed in the text.
 }
 \label{fig,loop1}
 \end{figure}
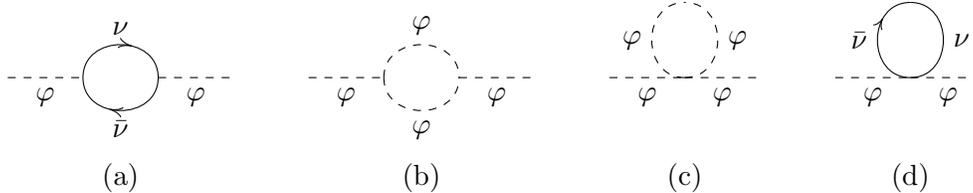

\section{Cosmological Evolution}
\label{sec:cosmo}

In the previous sections we have discussed various theoretical and observational aspects of our neutrino-assisted early dark energy model, and have derived several conditions for it to constitute a radiatively stable low energy effective field theory. In this section, we will apply the results of this study to elucidate the cosmological evolution of this model from early (the pre-recombination era) to late times. We begin by summarizing the salient results of the discussion above:
\begin{itemize}
    \item Treating our model as a low energy effective field theory mandates that we include a mass term in the bare potential. Once present, the mass receives corrections from neutrino loops that arise from the conformal coupling. The mass $m$ is radiatively stable to these corrections provided we take the neutrinos to be Dirac, and the mass to satisfy
    \begin{equation}\label{eq:qc2}
    m_\text{bare} \gsim\frac{\beta}{2\pi}\left(\frac{ m_\nu}{\mpl}\right)m_\nu\sim 1.5\times 10^{-27}\left(\frac{\beta}{100}\right)\left(\frac{m_\nu}{0.3 \textrm{eV}}\right)^2\textrm{eV}.
    \end{equation}
    \item If the scalar-neutrino coupling is linear, the minimum of the effective potential grows towards increasing $|\phi|$ in the past, and the field will exit the regime of validity of the effective field theory at some high redshift. This problem can be avoided if the higher order terms in the coupling resum into a function $A(\phi)$ that has a minimum at $\phi=-\bar\phi$ ($\bar\phi>0$) with $\beta\bar\phi<\mpl$. In this case $\phi\rightarrow\bar\phi$ at early times and the theory never exits the regime of validity of the EFT.
\end{itemize}

\subsection{Qualitative Overview of the Cosmic Evolution}
With the above considerations, a qualitative description of the evolution of the field is as follows. For convenience, we repeat the effective potential: 
\begin{equation}
    V_{\rm eff}(\phi)=\frac12m^2\phi^2+\frac{\lambda}{4} {\phi^4}+(\rho_\nu-3P_\nu)\ln\left[ A(\phi)\right].
\end{equation}
At high redshifts, $(\rho_\nu-3P_\nu)\gg m^2\phi^2,\, \lambda\phi^4$ so the effective potential is minimized by $\phi=-\bar\phi$, which minimizes $\ln A(\phi)$. The field is therefore fixed motionless at this minimum. From this we can derive the initial conditions $\phi_{\rm in}=\bar\phi$, $\dot\phi_{\rm{in}}=0$. It is worth remarking that, from a data analysis perspective, this scenario is more appealing than one where the field continually evolves at early times. In the latter scenario both the initial field value and the initial value of $\dot\phi$ are free parameters of the model, whereas in the former scenario the sole free parameter is $\bar\phi$. Reducing the number of free model parameters is likely to improve the goodness of fit.

As the universe expands and $(3P_\nu-\rho_\nu)$ redshifts, the contribution from the quartic part of the scalar potential is no longer negligible, and the minimum begins to evolve adiabatically away from $\bar\phi$ towards zero. This happens when
\begin{equation}
    \lambda{\bar{\phi}}^3\sim \frac{\beta}{\mpl}\left(3P_\nu-\rho_\nu\right).
\end{equation}
At this point, the higher-order terms in the expansion of the coupling function are no longer relevant. The evolution proceeds uninterrupted until $T_\nu\sim m_\nu$, at which point the neutrinos become non-relativistic and inject energy into the scalar, resulting in the kick feature in $\Omega_\phi$ discussed above. This signals the onset of the early dark energy phase, responsible for resolving the Hubble tension. The extra component of early dark energy causes the Hubble constant to redshift at a slower rate than that predicted by $\Lambda$CDM, lowering the sound horizon, and increasing the value of $H_0$ inferred from the CMB (see the discussion in the introduction or \cite{Knox:2019rjx,Sakstein:2019fmf} for more details on this). The energy injection is short lived so the kick in $\Omega_\phi$ is only a transient one, after which the field reverts to tracking the minimum of the effective potential, which continues to decrease towards zero as the neutrino density redshifts. This implies that at some point the quartic term will become negligible so that the mass term dominates, and the minimum of the effective potential is
\begin{equation}
\label{eq,phiminmatter}
    \phi_{\rm min}=-\left(\frac{\beta\rho_\nu}{\mpl m^2}\right),
\end{equation}
where we have taken $P_\nu\ll\rho_\nu$ corresponding to $T_\nu\ll m_\nu$. 

During the matter dominated era, the field can be decomposed into an adiabatic component that tracks $\phi_{\rm min}$, and a rapidly-varying component $\delta \phi$ corresponding to oscillations about the (time-dependent) minimum. The adiabatic component acts as an additional dark energy component, so that the energy density parameter of dark energy (including the cosmological constant) is
\begin{equation}
\rho_{\rm DE}\approx \Lambda\mpl^2+  \frac12m^2\phi_{\rm min}^2(t).
\end{equation}
Using Equation~\eqref{eq,phiminmatter}, one finds that the correction to the cosmological constant is 
\begin{equation}
    \frac{\rho_{\rm DE}}{\rho_{\Lambda}}=1+3\beta^2\frac{\Omega_{\nu,0}^2}{\Omega_{\Lambda,0}}\frac{H_0^2}{m^2}\approx 1+1.5\times10^{-12}\beta^2\frac{\Omega_{\nu,0}^2}{\Omega_{\Lambda,0}}\left(\frac{10^{-27}\textrm{ eV}}{m}\right)^2.
\end{equation}
Therefore, the values of the parameters necessary for our model to be relevant to the Hubble tension imply that dark energy will be driven almost entirely by the cosmological constant, and that the scalar will add a negligible contribution to its energy density. This implies that the scalar will lead to a small time-variation of the equation of state of dark energy, deviating slightly from $\omega_{\rm DE}=-1$. To estimate this, we demand that the total dark energy satisfy a continuity equation of the form 
\begin{equation}\label{eq:fakecont}
    \dot\rho_{\rm DE}+3H(\rho_{\rm DE}+P_{\rm DE})=0.
\end{equation}
Strictly speaking, the energy density of the scalar field, $\rho_\phi=\dot\phi^2/2+V(\phi)$, does not satisfy a continuity equation of this form due to the coupling to neutrinos. However, the purpose of the definition above is to connect our model with phenomenological parameterizations of the equation of state of dark energy e.g. $\omega$CDM or $\omega_0$--$\omega_a$. Observational analyses typically report bounds on these parameterizations rather than fitting to individual model. Using the definition \eqref{eq:fakecont}, we find that our model fits into the $\omega$CDM parameterization, where dark energy is driven by a fluid with constant equation of state $\omega_{\rm DE}$. We find
\begin{equation}
    \label{eq:eosDE}
    \omega_{\rm DE}=\frac{P_{\rm DE}}{\rho_{\rm DE}} = -1 -\frac{m^2\phi_{\rm min}\dot\phi_{\rm min}}{3H\mpl^2\Lambda}=-1+3\times10^{-12}\beta^2\frac{\Omega_{\nu,0}^2}{\Omega_{\Lambda,0}}\left(\frac{10^{-27}\textrm{ eV}}{m}\right)^2 \ ,
\end{equation}
at late times. We expect the data to prefer as small a mass as possible, which implies $m$ should be close to its one-loop quantum correction for reasons of radiative stability. Thus, Equation~\eqref{eq:eosDE} implies that the correction to $w_{\rm DE}=-1$ is negligible.

The rapidly-varying component, $\delta\phi$, of $\phi$ exhibits oscillations around the nearly quadratic potential
\begin{equation}
    V(\delta\phi) = \frac{\lambda}{4} \delta \phi^4 +\lambda \delta \phi^3 \phi_{\rm min}+\frac{3\lambda}{2}\delta \phi^2\phi_{\rm min}^2+ \frac{m^2}{2}\delta \phi^2.  
\end{equation} 
The cross terms and the quartic term are negligible, so the equation of motion is
\begin{equation}
\label{eq,deltaphieom}
    \ddot{\delta\phi}+3 H \dot{\delta\phi} +m^{2} \delta\phi=0 \ ,
\end{equation}
with solution  $\delta\phi = \frac{\delta\phi_0}{a^{3/2}} \cos(m t + \alpha)$, and corresponding fractional energy 
\begin{equation}
    \Omega_{\delta \phi} \equiv \frac{\frac{1}{2}\dot{\delta\phi}^2 + V(\delta\phi)}{3H^2\mpl^2}
\end{equation}
decaying as $a^{-3}$. This component therefore  corresponds to an additional contribution to the matter density, which can be a potential source of tension with the data, and is the motivation behind neglecting a scalar mass term in earlier EDE models. It is an important question whether EDE models can include a mass that is large enough that they are radiatively stable, but small enough that they are not in tension with the data. 
\begin{figure}
    \centering
    \includegraphics[width=0.85\textwidth]{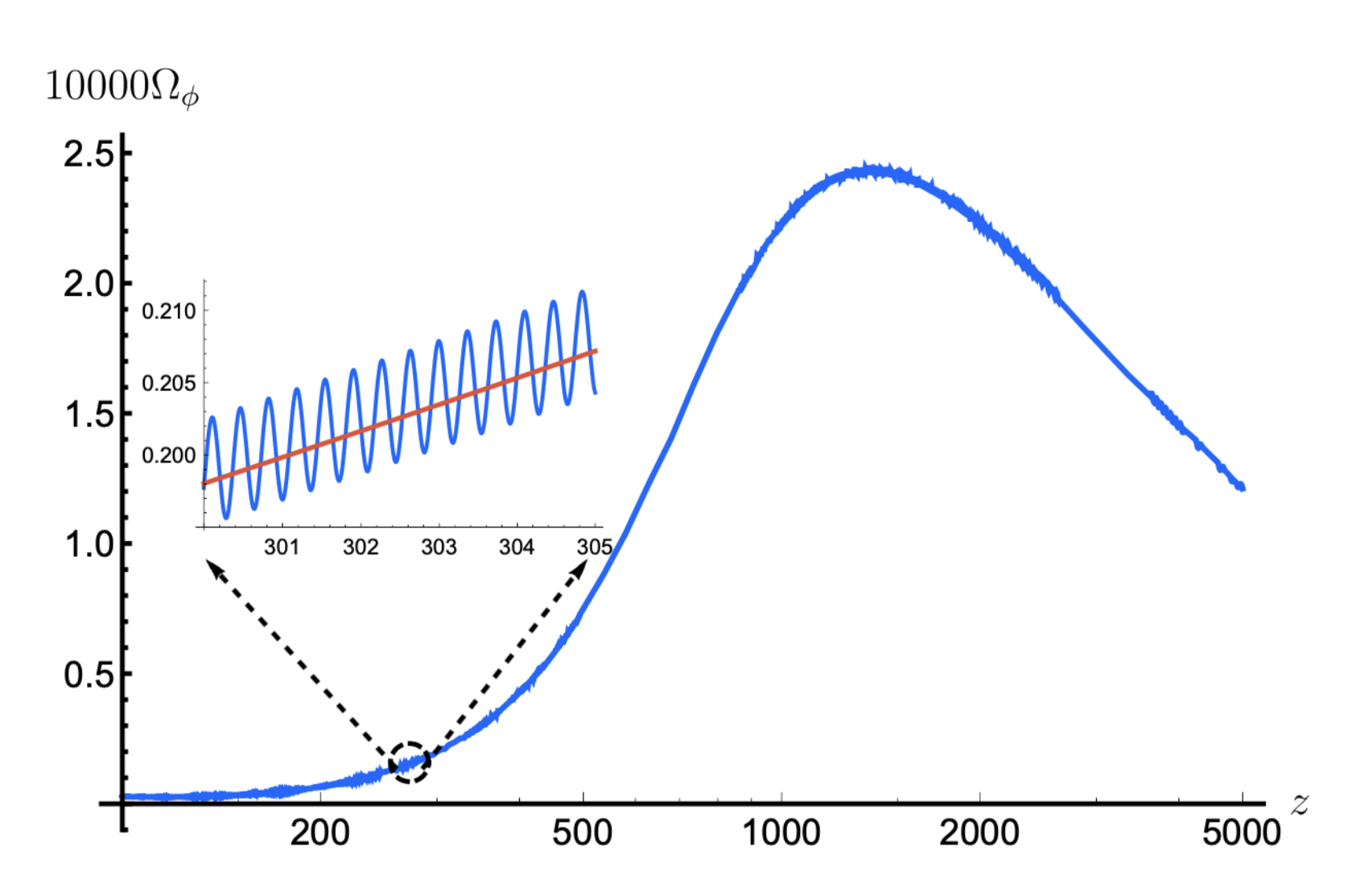}
    \caption{The evolution of $\Omega_\phi$ as a function of redshift. The specific parameter values are $m= \sqrt{30} \, \delta m = 1.63\times 10^{-26}\, \rm{eV}$ ($\delta m$ is the one-loop quantum correction to the mass defined in Equation~\eqref{qmass}), $\lambda= 10^{-98}$,  $\beta= 500$, and $m_\nu= 0.3\, \rm{eV}$. The zoomed in section shows the expected late-time harmonic oscillations due to the mass of the scalar field becoming the dominant term in the potential. The red line in the zoomed in region corresponds to  $\Omega_{\phi}(\phi = \phi_{\rm min} )$. }
    \label{fig:kicknum}
\end{figure}

\begin{figure}
    \centering
     \includegraphics[width=0.43\textwidth]{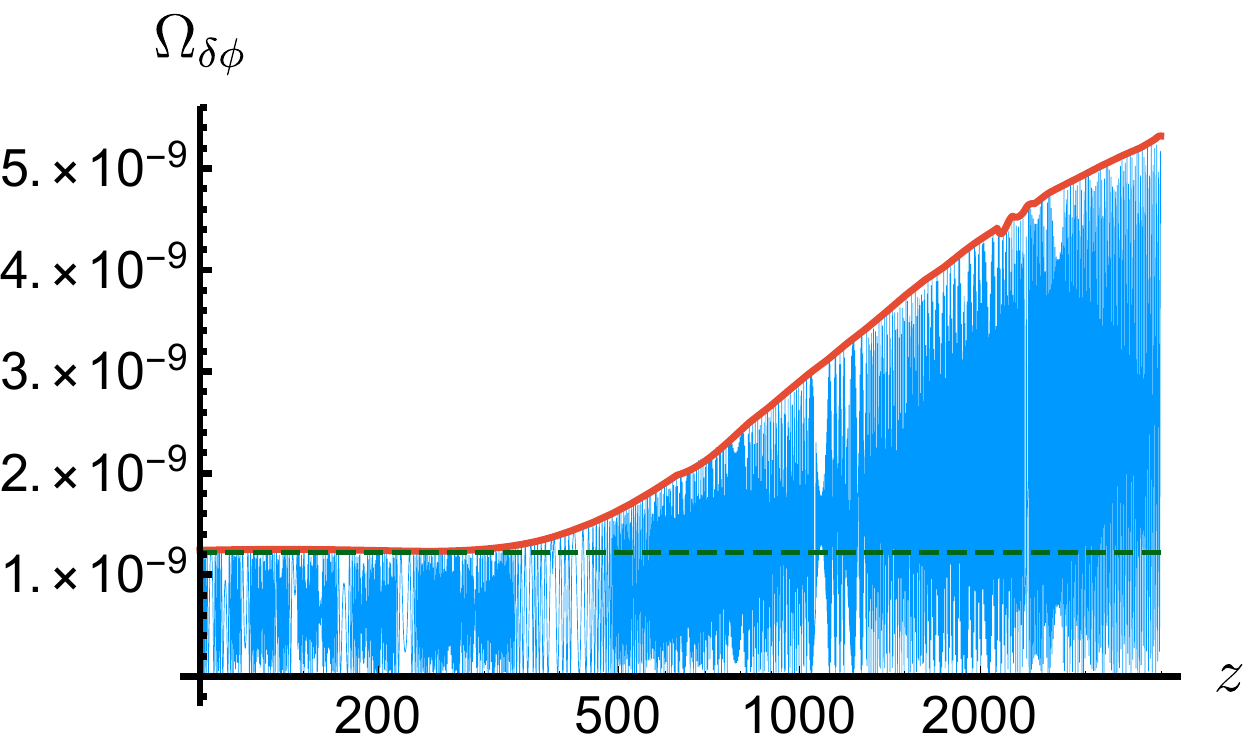}
       \includegraphics[width=0.43\textwidth]{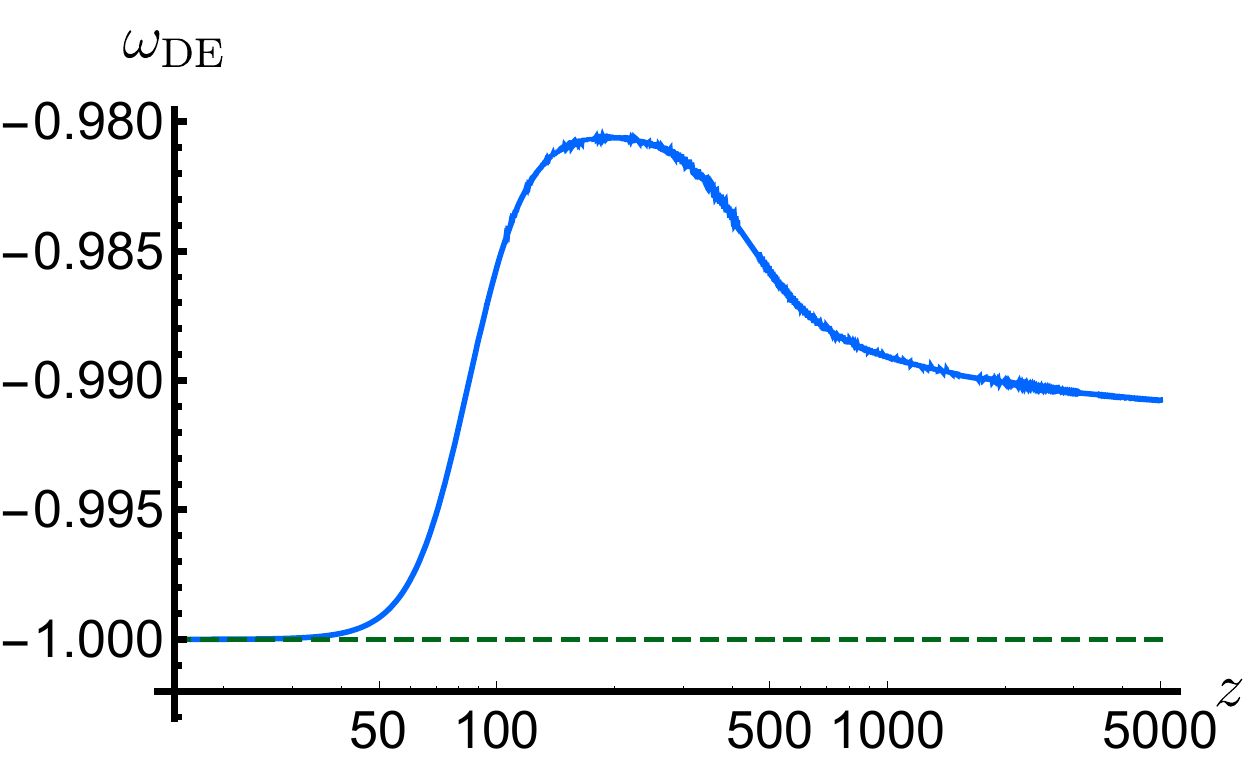}
    \caption{{\bf Left:} The evolution of $\Omega_{\delta\phi}$ as a function of redshift. The value of the scalar mass was chosen to be $m^2 = 5 \delta m^2 $. {\bf Right:} The effective equation of state of dark energy, including the contribution from the slowly varying component of the field $\phi_{\rm min}$. In both panels, the parameters were chosen to be $\beta = 800,~ \lambda = 10^{-98},~ m_\nu = 0.3 $ eV. 
    }
    \label{fig:masseffect}
\end{figure}

\subsection{Numerical Results}

We have numerically solved the coupled Friedman-neutrino-scalar system of equations, Eq.~\eqref{eq:fermieos} and Eq.~\eqref{eq:eom1}, including the other components of the energy budget of the universe, to quantitatively verify the qualitative expectations above, and to explore the parameter space. A representative example is shown in Figure~\ref{fig:kicknum}. All of the qualitative features discussed above are evident. Namely, the kick happens around $\mathcal{O}(z)\sim 1000$, and at late times we have two components, a slowly varying component, $\phi_{\rm min}$, and an oscillating component,  $\delta\phi$. 
Note that this model is most likely to help solve the $H_0$ tension when the kick magnitude is the largest. 
Therefore, one aim of this work is to perform a complete exploration of the parameter space to find such regions and inform follow-up studies attempting to constrain our model using cosmological data sets. 

The novel effects of the mass term are exemplified in the left-hand side of Figure~\ref{fig:masseffect}, where we plot the evolution of the density parameter for the oscillating component of the field $\delta\phi$. One can see that this component of the field exhibits anharmonic oscillations at early times that harmonize during the matter era so that $\Omega_{\delta\phi}(z)$ tends to a constant, indicating that it is acting as an additional component of dark matter. The right hand panel of Figure~\ref{fig:masseffect}, shows the evolution of the equation of state of dark energy due to the slowly-varying component $\phi_{\rm min}$ which is given by Equation~\eqref{eq:eosDE}. One can see that deviations from $w_{\rm DE} = -1$ are indeed negligible at late time. 

We have performed a thorough exploration of the parameter space in order to explore the correlation between the magnitude of the kick and the model parameters. The relevant parameters are the mass of the scalar field $m$, the quartic self-coupling $\lambda$, and the coupling to neutrinos $\beta$. The evolution of the universe at matter-radiation equality, and at later times is independent of $\bar\phi$,  so we will not explore this parameter. We also allow the neutrino mass $m_\nu$ to vary modestly from the best-fit $\Lambda$CDM value obtained from Planck data. We consider this reasonable at this stage, since the reported bounds are found from the Planck posterior, and are thus likely to change when additional parameters are added to the model, and the cosmology is altered. In particular, the Boltzmann hierarchy for massive neutrinos will be modified due to the non-minimal coupling to the scalar \cite{Oldengott:2014qra}.  We are particularly interested in regions of parameter space where the kick magnitude is the largest around the epoch of matter-radiation equality. We will not attempt to constrain the model using cosmological data sets in this work; its focus being on theoretical model building and phenomenology. For this reason, we will also not make any speculative conclusions regarding which regions of parameter space do or do not resolve the Hubble tension, postponing this for future work that will use data analysis to answer this question definitively. The regions of parameter space where we find the largest kick magnitude are shown in Figure~\ref{fig,kickcontour2}. 

\begin{figure}
   \centering
    \includegraphics[width=0.43\textwidth]{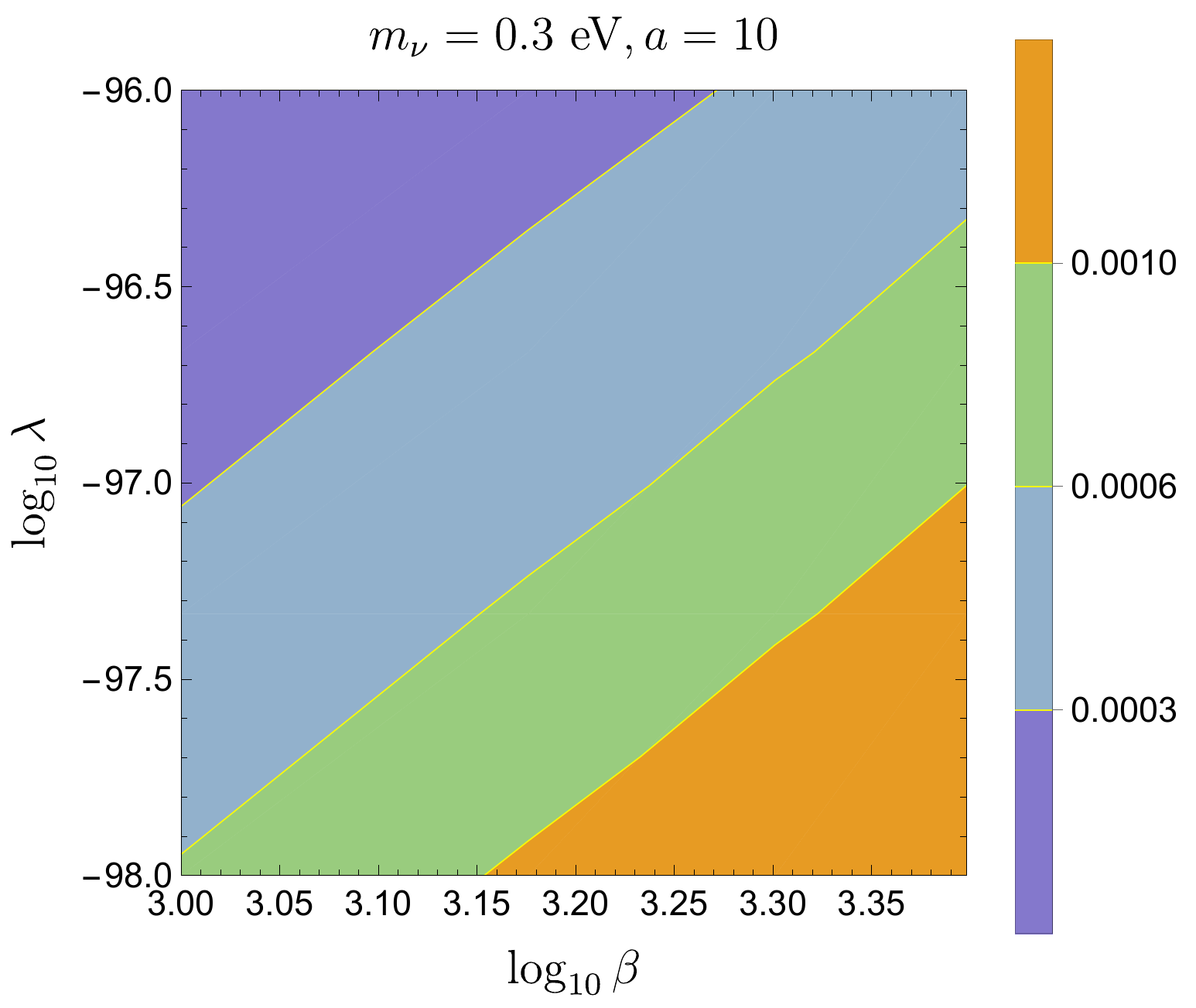}
    \includegraphics[width=0.43\textwidth]{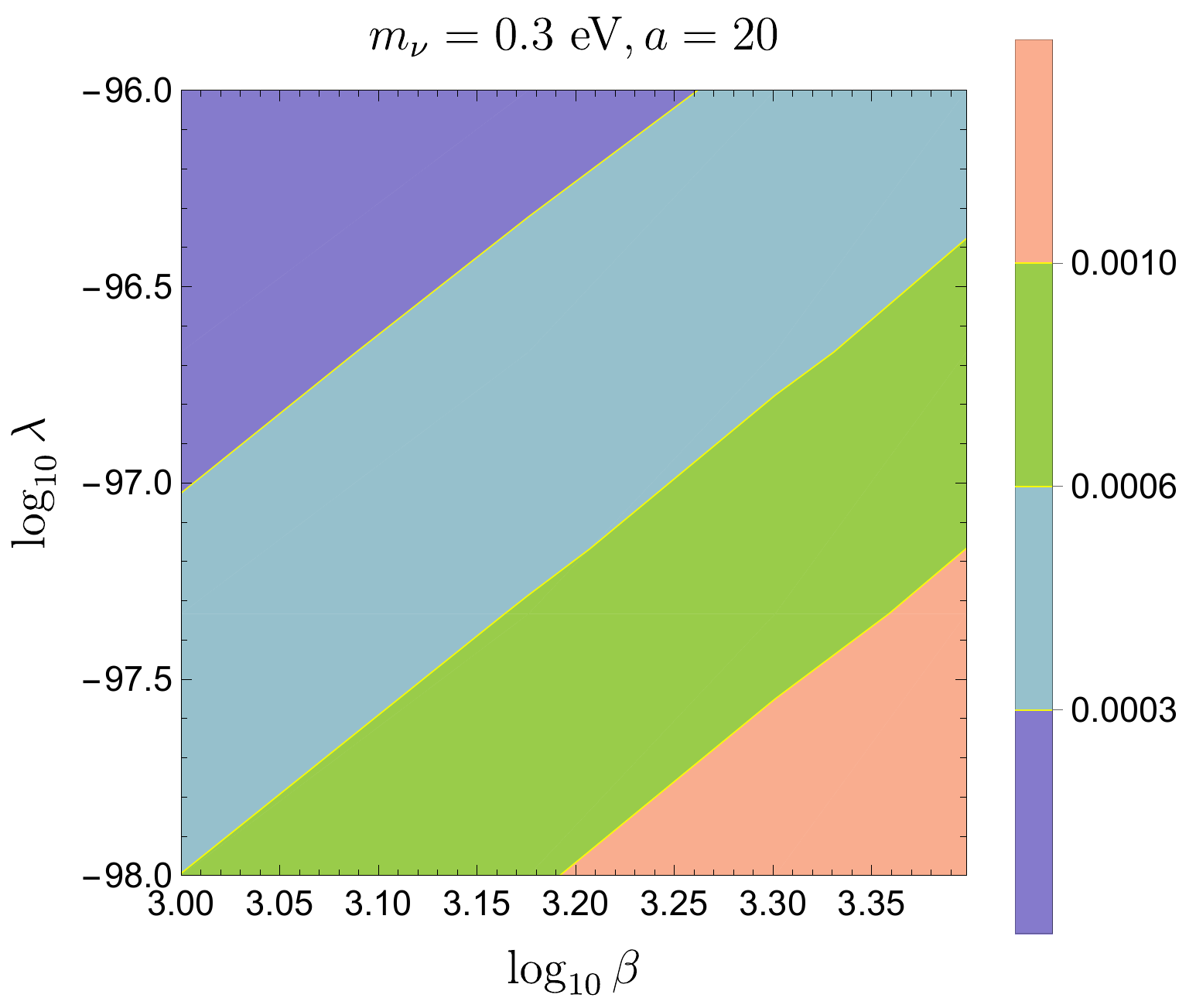}
        \includegraphics[width=0.43\textwidth]{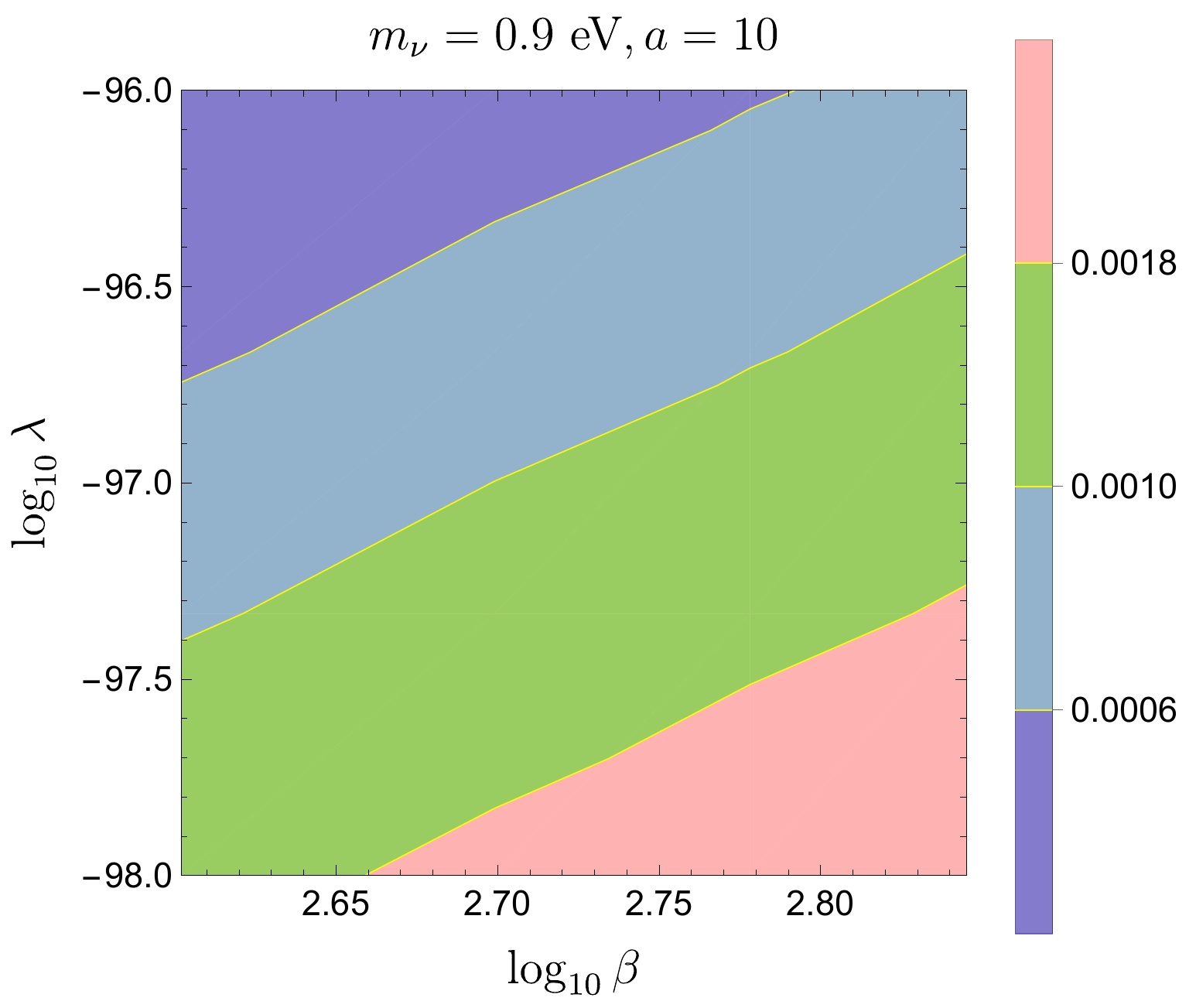}
 \caption{Contour plot of the kick's magnitude as a function of $\log_{10}\beta$ and $\log_{10}\lambda$ for different values of $m_\nu$ and different $m^2 = a \delta m^2$ (see Equation~\eqref{qmass}) indicated in the figure. The magnitude of the kicks is slightly smaller for larger $a$.
 } \label{fig,kickcontour2}
 \end{figure}

\section{Discussion and Conclusions}
\label{sec:concs}

In this work, we have investigated theoretical and phenomenological aspects of the recently proposed neutrino-assisted early dark energy resolution of the Hubble tension. To this end, building on the simplest toy model, we have constructed a more theoretically robust model that satisfies the requirements of effective field theory and radiative stability, and have explored its cosmological phenomenology. In this final section, we discuss potential model extensions, survey the state of the field, and draw our conclusions.

\subsection{Observational Probes} 
\label{sec:constraints}

In this section we explore potential observational signatures of our model that result from the scalar-neutrino coupling rather than from the modified cosmological dynamics. These represent complementary novel probes of our neutrino-assisted dark energy scenario.

\subsubsection{Constraints on neutrino self-interactions}

Constraints on neutrino self-interactions mediated by a light scalar field have been analyzed in \cite{Forastieri:2015paa,Forastieri:2019cuf}. The neutrino self-interactions alter the cosmological evolution and thus affect the CMB. Using Planck data, the strengths of such interactions can be constrained. These current bounds cannot be directly applied to our model, since the cosmological evolution is different in our case. Instead, we summarize their results here, noting that once the model presented here is implemented into cosmological solvers and MCMC codes, a similar procedure could be applied.

Consider neutrino-neutrino scattering mediated by a light scalar in the relativistic limit $x\gg1$. The cross-section scales as
\begin{equation}
\sigma\sim \frac{\beta^4 m_\nu^4}{\mpl^4 s}  \ ,
\end{equation}
where $s$ is the usual Mandelstam variable. The thermally averaged cross-section then reads
\begin{equation}
\langle\sigma v \rangle \sim \frac{\beta^4 m_\nu^4}{\mpl^4 T_\gamma^2} \ ,
\end{equation}
which leads to a scattering rate $\Gamma\sim n_\nu \langle\sigma v \rangle $ that grows as $T_\gamma$. Since the scattering rate decreases slower than the expansion of the Universe, neutrino collisions can become relevant even after the neutrinos decouple at $T_\gamma\sim 1 $ MeV. The constraint obtained in \cite{Forastieri:2019cuf} is given for the neutrino-scalar Yukawa coupling, which in our case is $y_{\phi  \bar{\nu}\nu}=\beta m_\nu / \mpl\lesssim 10^{-7}$, 
and is valid for $m_\nu\leq0.06$ eV. This leads to a very weak upper limit on $\beta$
\begin{equation}
\beta\lesssim \left(\frac{10^{-2} \text{eV}}{m_\nu}\right)  10^{23} \ .
\end{equation}
However, we reiterate that an updated analysis based on the modified cosmology of our model is necessary to obtain a definitive bound on $\beta$.

\subsection{Yukawa Potential and Weak Equivalence Principle Violations}
\label{sec:WEP}

The interaction between neutrinos and the scalar leads to a fifth force between neutrinos with magnitude a factor of $2 \beta^2$ times the strength of their gravitational attraction. For a massless field, the potential is the usual Newtonian one ($\propto1/r$), but when the scalar is massive there is a Yukawa suppression. In our case, the potential between the neutrinos generated due to the $\phi\nu\bar\nu$ interaction is 
\begin{equation}
V^{\bar{\nu}\nu}_\phi(r)=-2 \beta^2\frac{Gm_\nu^2}{r}e^{-mr} .
\end{equation}
At distances $r\ll 1/ m $, the potential is not suppressed. This shows that at small distances the neutrinos feel a fifth force which behaves as a correction to the gravitational force 
\begin{equation}
V^{\bar{\nu}\nu}_\text{eff}=V^{\bar{\nu}\nu}_{G}+V^{\bar{\nu}\nu}_{\phi}\simeq -(1+2\beta^2)  \frac{Gm_\nu^2}{r} \  ,
\end{equation}
from which we can define an effective gravitational coupling
\begin{equation}
G^\text{eff}_N=(1+2\beta^2) G_N. \label{geff}
\end{equation}
This phenomenon is just a manifestation of the violation of the Weak Equivalence Principle (WEP) in the neutrino sector that arises because we have chosen to couple the EDE scalar to neutrinos solely, and not other matter species. The violation arises because neutrinos follow the geodesics of the metric $\tilde{g}$ while the rest of the standard model follows the geodesics of $g$. 

Cosmologically, when $m_\nu\gg T_\nu$ i.e. at late times, the perturbation of the neutrinos density in the Newtonian gauge evolves according to \cite{Brax:2004qh,Brax:2005ew,Brax:2012gr,Brax:2013yja}
\begin{equation}
   \ddot \delta_\nu + 2H\dot\delta_\nu-\frac{3}{2}\Omega_\nu(a)\left(1+\frac{\Delta G_\nu(k,\beta,m)}{G_{\rm N}}\right)\delta_\nu=0,
\end{equation}
where
\begin{equation}
    \frac{\Delta G_\nu(k,\beta,m)}{G_{\rm N}} = \frac{2\beta^2}{1+\frac{m^2a^2}{k^2}}.
\end{equation}
On scales $k\ll m$, and taking $a\sim1$ since we are interested in late times, we have
\begin{equation}
    \frac{\Delta G_\nu(k,\beta,m)}{G_{\rm N}} \approx  1 \ \left(\frac{\beta}{10^{3}}\right)^2 \left(\frac{k}{(6.4 \textrm{ Mpc})^{-1}}\right)^2\left(\frac{10^{-27}\textrm{ eV}}{m}\right)^2
\end{equation}
whereas for $k\gg m$ one finds $\Delta G_\nu(k,\beta,m)/G_{\rm N}=2\beta^2$. Currently, there are no experimental constraints that result from either the Yukawa force or the weak equivalence principle violation, but it may be possible to place future bounds by looking for effects of the fifth force on cosmological observables such as the non-linear matter power spectrum \cite{Bird:2011rb}, and the clustering properties of neutrinos inside voids \cite{Massara:2015msa,Schuster:2019hyl}, both of which should be observable with the next generation of lensing surveys \cite{Banerjee:2019omr}.

\subsection{Model Extensions}

\subsubsection{Couplings to Other Standard Model Particles and the Dark Sector}

The minimal model only required us to couple the scalar to neutrinos, but such a sequestering is likely unnatural since the requirements of the standard model gauge group requires the left-handed neutrinos to transform as an $\mathrm{SU(2)}$ doublet. From an EFT perspective, one would then expect couplings to other standard model particles. Similarly, the neutrino mass term we have considered breaks $\mathrm{SU(2)}$, implying that a UV-completion is needed to embed our model into the standard model, and thus requiring that scalar-Higgs couplings should be present in the EFT. If the neutrinos acquire their mass through a coupling to a dark sector Higgs, scalar-dark matter couplings are also be expected.

In general, we expect that the scalar field will couple to different particle species $i$ with different strengths $\beta_i$. This would lead to violations of the weak equivalence principle (WEP) in the visible and dark sectors. The WEP in the dark sector can be tested by looking for the effects on the tidal disruption of Milky Way Satellites \cite{Kesden:2006vz,Kesden:2006zb}, yielding $\beta_{\rm DM}<\mathcal{O}(0.1)$, and, recently, tests of the WEP between baryons and dark matter using the predicted warping of galaxies has constrained $\beta_{\rm DM}^2-\beta_{\rm SM}^2<5\times10^{-5}$ \cite{Desmond:2018kdn,Desmond:2020gzn}. Violations of the WEP within the standard model are more tightly constrained  \cite{Will:2014kxa,Sakstein:2017pqi,Sakstein:2017bws,Bartlett:2020tjd}. If the coupling to matter is universal, these additional bounds must be applied to $\beta$. If the coupling is non-universal, they need only apply to $\beta_i$ where $i$ is the matter species relevant for the specific probe in question.

Additional conformal couplings imply additional Yukawa interactions between the scalar and fermions of the form $y_{\phi\bar\psi_i\psi_i}\phi\bar\psi\psi$ where $y_{\phi\bar\psi_i\psi_i}=\beta_i m_\psi/\mpl$. Since our coupling is Planck suppressed, the largest Yukawa couplings will arise for the heaviest particles. For example, the Yukawa coupling for the top quark would be
\begin{equation}
y_{\phi  \bar{t}t}=\beta_t\frac{ m_t}{M_{Pl}} = \left(\frac{\beta_t}{0.1}\right)  10^{-18} \ .
\end{equation} 
We expect $\beta_i\le\mathcal{O}(1)$, and it is therefore unlikely that these couplings could be probed using terrestrial colliders.

Another consequence of additional couplings is additional quantum corrections to the mass of the scalar field from fermions. Following the discussion in section \ref{sec:quantum}, if the coupling of the scalar to a fermion $i$ with mass $m_i$ (dark or standard model) is $\beta_i$ where $\mathcal{O}(\beta_i) \approx \mathcal{O}(\beta)$, the EDE's mass receives a quantum correction of order of $\delta m(\beta_i)=\beta_i^2m_i^4/\mpl^2 \gg \beta^2m_\nu^4/\mpl^2 $. In order to render the model radiatively stable to these corrections, one must either take the mass of the scalar to be heavier than the largest quantum correction, or tune $\beta_i$ such that these contributions are sub-dominant to the neutrino-induced radiative corrections. In the former case, one may run up against constraints from late time probes of the background cosmological expansion. In the latter case, the required tunings are likely unnatural and one must check that they themselves are not destabilized by quantum corrections. If the coupling to matter is universal, the largest contribution to the quantum correction to the mass of the scalar field is due to the heaviest particle, i.e. the top quark, which is of order $10^{-6}\beta_t$ eV. Additional couplings also imply additional kicks in the early Universe. These are not relevant for the Hubble tension.

Lastly, we briefly comment on a coupling that could arise naturally in a UV completion of this model; this is a coupling to the Higgs field of the form $A(\phi)\lambda_h\bar{L}H\nu_R$, where $L$ is a lepton doublet, $H$ is the Higgs scalar doublet, and $\nu_R$ a right-handed neutrino. After spontaneous symmetry breaking this leads to a coupling $A(\phi)\lambda_h  h \bar{\nu}\nu$, where we can now identify $\lambda_h=\sqrt{2}m_\nu/v$, with $v$ the Higgs vacuum expectation value. The leading order interaction with the EDE field is then given by $(\sqrt{2} \beta m_\nu/v \mpl) \phi h \bar{\nu}\nu=\phi h\bar\nu\nu/\mathcal{M}$ with $\mathcal{M}\sim 10^{26}/\beta$ TeV, implying that that the corrections to Higgs processes are not observable, nor will they be in the near future. This follows from the fact that such interactions are Planck suppressed. 

\subsubsection{Couplings to Different Neutrino Mass Eigenstates}
Consider an action which involves all the standard model neutrinos, where each of the mass eigenstates is conformally coupled to $\phi$ with a different strength $\beta_i$. This would arise by generalizing the metric used to contract indices to $\tilde{g}^{(i)}_{\mu\nu}=A^2_i(\phi)g_{\mu\nu}$. The neutrino sector is now given by
\begin{equation}
S_\nu=\int d^{4} x \sqrt{-g}\sum_{i=1}^3\left(i \bar{\nu_i} \gamma^{\mu} {\nabla}_{\mu} {\nu_i}
-\frac{\beta_i m_{\nu} }{\mpl}\varphi\bar{\nu_i} {\nu_i}\right) \ ,
\end{equation}
with 
\begin{equation}
\beta_i=\mpl \, \frac{\dd \log{A_i}(\phi)}{\dd\phi}\Bigg|_{\phi=\phi_\text{min}(z=0)} \ ,
\end{equation}
where the label $i$ denotes different neutrino mass eigenstates. To obtain these expressions, we expanded the field $\phi$ around its minimum as $\phi=\phi_\text{min}(z=0)+\varphi$, performed a Taylor expansion, and neglected higher orders in $\varphi/\mpl$. We can see that there is a different effective gravitational coupling, of the form of Equation~\eqref{geff}, for each neutrino mass eigenstate. Similar violations of the equivalence principle in the neutrino sector have been proposed previously, since they lead to changes to the details of neutrino oscillations \cite{Gasperini:1988zf,Gasperini:1989rt,Halprin:1991gs,Yasuda:1994nu,Adunas:2000zm,Diaz:2020aax,GonzalezGarcia:2004wg,GonzalezGarcia:2007ib}. The probability of oscillations depends on the difference in energies between the neutrinos involved. In the presence of violations of the WEP we have \cite{Diaz:2020aax},
\begin{equation}
    \Delta E_{i j}=\frac{\Delta m^{2}}{2 E}+4 E  \Phi \, \Delta\beta^2_{i j} \ ,
\end{equation}
where $\Phi$ is the Newtonian potential and $\Delta\beta^2_{ij}\equiv \beta^2_i-\beta^2_j$ , with $i, j$ labeling different mass eigenstates. The first term leads to the usual oscillations due to the different mass eigenstates and the second arises from the different gravitational couplings between the neutrinos. Current bounds from combining Super–Kamiokande and the KEK to Kamioka long-baseline neutrino oscillation experiment (K2K) data give \cite{GonzalezGarcia:2004wg}
\begin{equation}
    |2 \Phi \Delta \beta^2_{ij}| \leq 4.0 \times 10^{-25} \ . \label{boundPhiBeta}
\end{equation}
Taking $\Phi\sim GM_\oplus/R_\oplus c^2\sim 10^{-9}$, where $M_\oplus$ and $R_\oplus$ are the Earth's mass and radius, one obtains a bound $\Delta \beta^2_{ij}\le 2\times 10^{-16}$. This implies that the EDE scalar should couple to all neutrino species with the same coupling strength to avoid fine-tunings. Future prospects using data from the DUNE experiment could improve this bound by an order of magnitude \cite{Diaz:2020aax}.

\subsection{Early Dark Energy as a Resolution of the Hubble Tension}

Recently, the success of the early dark energy paradigm for resolving the Hubble tension has been called into question by several works in the literature \cite{Hill:2020osr,Ivanov:2020ril,DAmico:2020ods}. They find that there is no strong preference for EDE over $\Lambda$CDM when large scale structure data is included in the analysis, and that the preference comes solely once low redshift data sets are included, in particular SH0ES. Furthermore, the $\sigma_8$-tension, which already exists within $\Lambda$CDM, is mildly exacerbated by including EDE. A recent counter-argument \cite{Smith:2020rxx} suggests that these conclusions are driven by the choice of priors, and by tensions that already exist with $\Lambda$CDM. 

We will not make any judgements about the validity of these claims or counter claims here, but will make two remarks on their relevance to our work, should they hold up to scrutiny. The first is that any analysis of cosmological datasets requires a model to be specified, and the conclusions may differ for competing EDE scenarios. For example, a recent analysis of the new early dark energy model \cite{Niedermann:2019olb,Niedermann:2020dwg} found that it was both compatible with large scale structure data and able to ameliorate the Hubble tension \cite{Niedermann:2020qbw}. Likewise, whether our model ultimately runs afoul of large scale structure surveys requires a separate analysis. The second remark is that the analyses performed so far apply strictly to the simplest scenario in which the EDE is an uncoupled scalar. Since our scalar is coupled to neutrinos, not only is the background cosmology modified, but also the details of structure formation through modifications to the neutrino free streaming length and Boltzmann hierarchy. Similarly, the potential extensions discussed above, especially a coupling to dark matter, will also lead to modifications to structure formation. Until these new effects are quantified, one must use caution in making inferences about our model based the results of the analyses discussed above.

\subsection{Conclusions}

The Hubble tension is one of the more tantalizing hints at new physics provided by the multitude of cosmological datasets currently available to us. As more data has accumulated over the last few years, the tension has remained, and while there may easily be systematic or other errors responsible, it is incumbent on theorists to explore what the discrepancy might mean should it persist. In this paper we have explored theoretical and phenomenological details of a solution to this problem originally proposed by two of us~\cite{Sakstein:2019fmf}. This proposal is based on a particularly fine-tuned proposal -- that of early dark energy. Early dark energy faces the problem that one must build in by hand the fact that it is active around the epoch of matter-radiation equality, since this is the period during which an injection of scalar field energy can alter the Hubble constant in such a way as to reconcile the two different measurements. Our proposal -- neutrino assisted early dark energy -- overcomes this by positing that the relevant scalar field couples to neutrinos, and then exploits the fact that neutrinos decouple at around the required epoch.

We have addressed three main issues in this paper. First, we have extended the proof-of-principle model put forward in our original paper to construct a well-behaved effective field theory, in which quantum effects are under control, and all operators not explicitly forbidden by symmetries are included. Second, we have performed a careful study of the cosmological evolution of this model. This required us to understand the initial condition challenges posed by the generic construction, and to propose a single extra simplifying assumption about the UV completion of the EFT that allows the model to remain within the domain of validity of the EFT. By understanding the background cosmology of this extended model in detail, and by studying the effects that the parameters of the theory have on the size of the neutrino ``kick", we have identified those parameter choices for which the theory has the best chance of providing a fully-consistent solution to the Hubble tension. These parameter choices will be an input into a future Markov Chain Monte Carlo (MCMC) comparison of the model to multiple datasets.

Finally, we have considered how the neutrino-assisted early dark energy model might be constrained by observations and experimental measurements. Even at the level of the basic model, it is possible to study the implications of a scalar-neutrino coupling for measurements of neutrino self-interactions and weak equivalence principle violations. {In the latter case
we have shown that there are potentially interesting hints that may be accessible to next
generation lensing surveys.}
Furthermore, we note that, while the model presented here is sufficient to understand the cosmological implications of this mechanism, a complete model would involve all allowed couplings to the standard model. This raises the possibility of ruling out or detecting the new interactions via various particle physics experiments. For the most part, we have confirmed that such tests do not strongly constrain our model due to the Planck-suppressed nature of the relevant couplings. However, extensions of the model in which the EDE field couples differently to different neutrino mass eigenstates have parts of their parameter space that are within the sensitivity of the upcoming DUNE experiment.

What remains for this model is a dedicated likelihood analysis to understand whether it is a viable solution to the Hubble tension. Existing analyses of EDE models do not apply to one in which the field is coupled to neutrinos, since the sizes of the allowed parameters can be quite different, and because the coupling modifies the details of structure formation as well as the background evolution of the universe. The results derived in this paper, as well as putting the model on a more sturdy theoretical footing, form the basis for such as MCMC approach, which we will carry out in a separate publication.

\section*{Acknowledgements}
We thank Eric Baxter, Bhuvnesh Jain, Tanvi Karwal, Danny Marfatia, and Marco Raveri for helpful comments and discussions. The work of QL and MT is supported in part by US Department of Energy (HEP) Award DE- SC0013528. M.T. is also supported by NASA ATP grant 80NSSC18K0694, and by the Simons Foundation Origins of the Universe Initiative, grant number 658904. MCG is supported by the STFC grant ST/T000791/1 and the European Union’s Horizon 2020 Research Council grant 724659 MassiveCosmo ERC–2016–COG.

\bibliography{ref}
\end{CJK*}
\end{document}